\documentclass[a4paper]{jpconf}
\usepackage{graphicx, subfigure, amssymb, cite}
\pdfoutput=1
\begin{document}
\title{Entropy and thermopower in the 2D t-J model}

\author{W O Putikka}

\address{Physics Department, The Ohio State University, 1760 University Dr, Mansfield, OH 44906, USA}

\ead{putikka.1@osu.edu}

\begin{abstract}
The entropy of the two-dimensional $t$-$J$ model is investigated using its 12th order high temperature series.  A direct
Pad\'{e} extrapolation of the entropy series doesn't converge well for temperatures below $T\sim J$.  The series coefficients
are exact polynomials so the series convergence can be improved by modifying the series that is extrapolated.  By subtracting
a scaled version of the series for the entropy of the Heisenberg antiferromagnet from the $t$-$J$ entropy series the 
low temperature
convergence is greatly improved.  Using this technique results are obtained for the full range of electron densities and
temperatures.  The electron density is an adjustable parameter in the series coefficients allowing the density dependence
of the entropy and the density derivative at fixed temperature $\partial S/\partial n|_T$ to be determined accurately.
The density derivative depends strongly on temperature, unlike noninteracting models.  The density derivative is also
an approximation to the experimentally measured thermopower.
\end{abstract}

\section{Introduction}

The two-dimensional $t$-$J$ model has been studied for many years as a model for the
copper oxide planes found in high temperature superconductors.  Despite considerable
effort\cite{prelovsek} the basic thermodynamic properties of this model are still not well understood.  
The calculation reported here investigates the entropy by means of its high temperature
series calculated to 12th order in inverse temperature $\beta$.  In principle, knowing
the full temperature dependence of the entropy and the value of the free energy at one
temperature is sufficient to fully determine the free energy and thus all the thermodynamic
properties of the model.  This more general calculation is left for a future publication.
In this paper our attention is restricted to the entropy per site $S$ and its density
derivative at fixed temperature $\partial S/\partial n|_T$.

The $t$-$J$ model Hamiltonian is given by
\begin{equation}
H = -t\sum_{\sigma\left<ij\right>}\left(c_{i\sigma}^{\dagger}c_{j\sigma}+c_{j\sigma}^{\dagger}c_{i\sigma}\right)
+J\sum_{\left<ij\right>}\vec{S}_i\cdot\vec{S}j,
\end{equation}
along with the constraint of no double occupancy.  The high temperature series for the 
Helmholtz free energy per site is calculated by a linked cluster expansion, taking the form
\begin{equation}
\beta F = \sum_{i=0}^N f_i (\beta J)^i,
\end{equation}
where the $f_i(n,t/J)$ are exact polynomials in terms of the electron density $n$
and the ratio of the model coupling constants $t/J$.  For the calculation reported here
$N=12$ and we fix $t/J=2.5$.  The series for the entropy per site is also calculated in terms of the $f_i$ as
\begin{equation}
S = k_{\rm B}\sum_{i=0}^N (i-1)f_i(\beta J)^i
\end{equation}
and the energy per site is
\begin{equation}
\beta E = \sum_{i=0}^N i f_i (\beta J)^i.
\end{equation}

\section{Details of the Calculation}

To reach low temperatures, the series expansion for the entropy needs to be extrapolated by
Pad\'{e} approximants.  With the available 12th order series the direct extrapolation of the $t$-$J$
entropy only converges for $T\gtrsim J$.  A low temperature scale remains unresolved in this calculation.
To improve the convergence we can try to extract this low 
temperature scale by forming the series for 
\begin{equation}\label{eq:deltas}
\Delta S(n,T) = S_{tJ}(n,T) - n_{AF}^*S_{AF}(J^*,T)-\alpha^* S_{TB}(n,T),
\end{equation}
where $S_{AF}$ is the entropy of the Heisenberg antiferromagnet (found by setting $n=1$ in the $t$-$J$ entropy),
$S_{TB}$ is the tight-binding model entropy and the starred parameters are adjusted to produce a function
$\Delta S$ with a simple temperature dependence that extrapolates well to low temperatures.
A least squares fit of the $[6/6]$ Pad\'{e} to a fourth order polynomial is calculated to extend $\Delta S$ to
$T=0$.  For $n\ge0.55$
a reasonable $\Delta S$ can be found by setting $n_{AF}^*=n$, $\alpha^*=0$ and adjusting $J^*$.  
\begin{figure}[h]
 \begin{center}
  \subfigure{\includegraphics[width=0.49\textwidth]{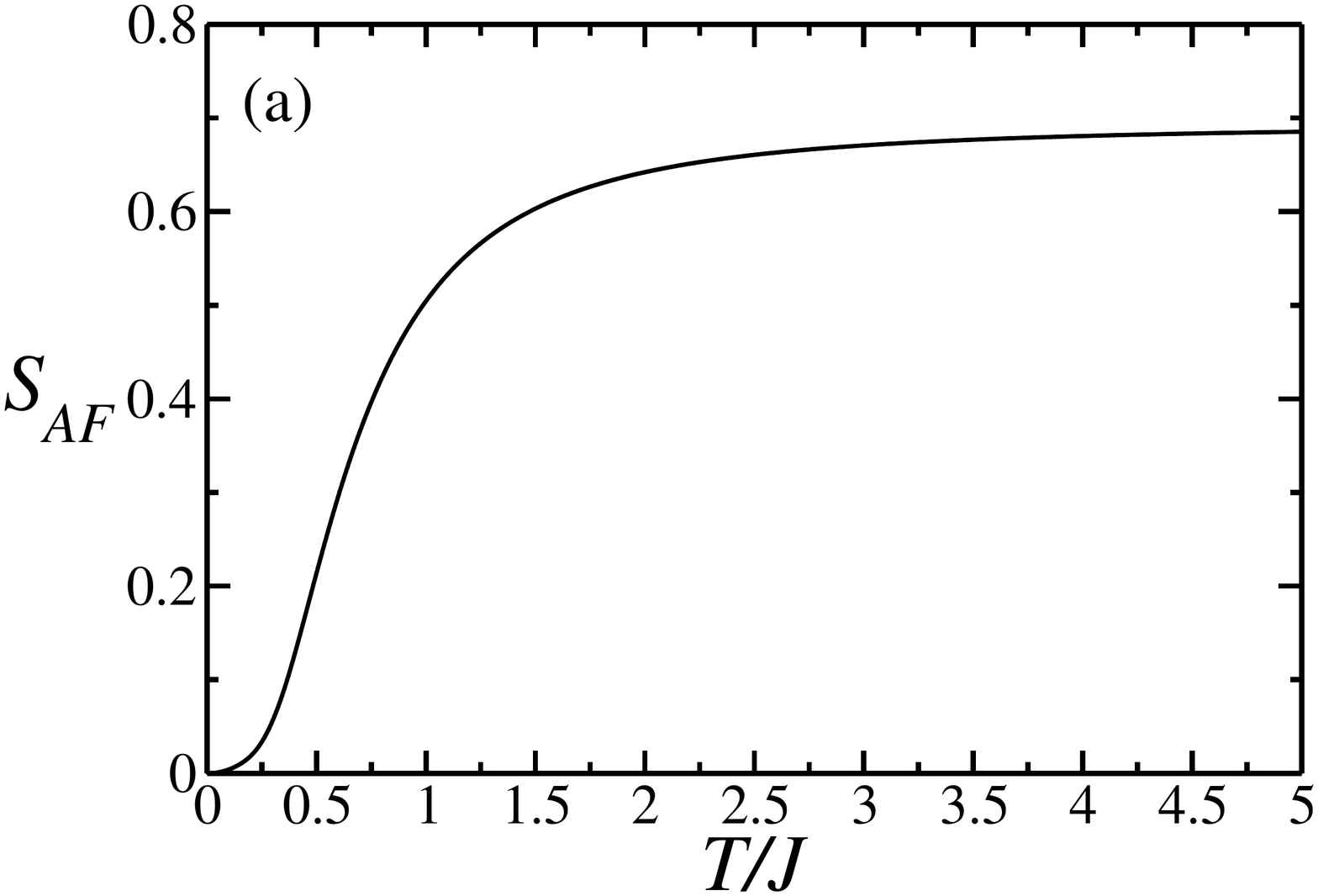}}
  \subfigure{\includegraphics[width=0.49\textwidth]{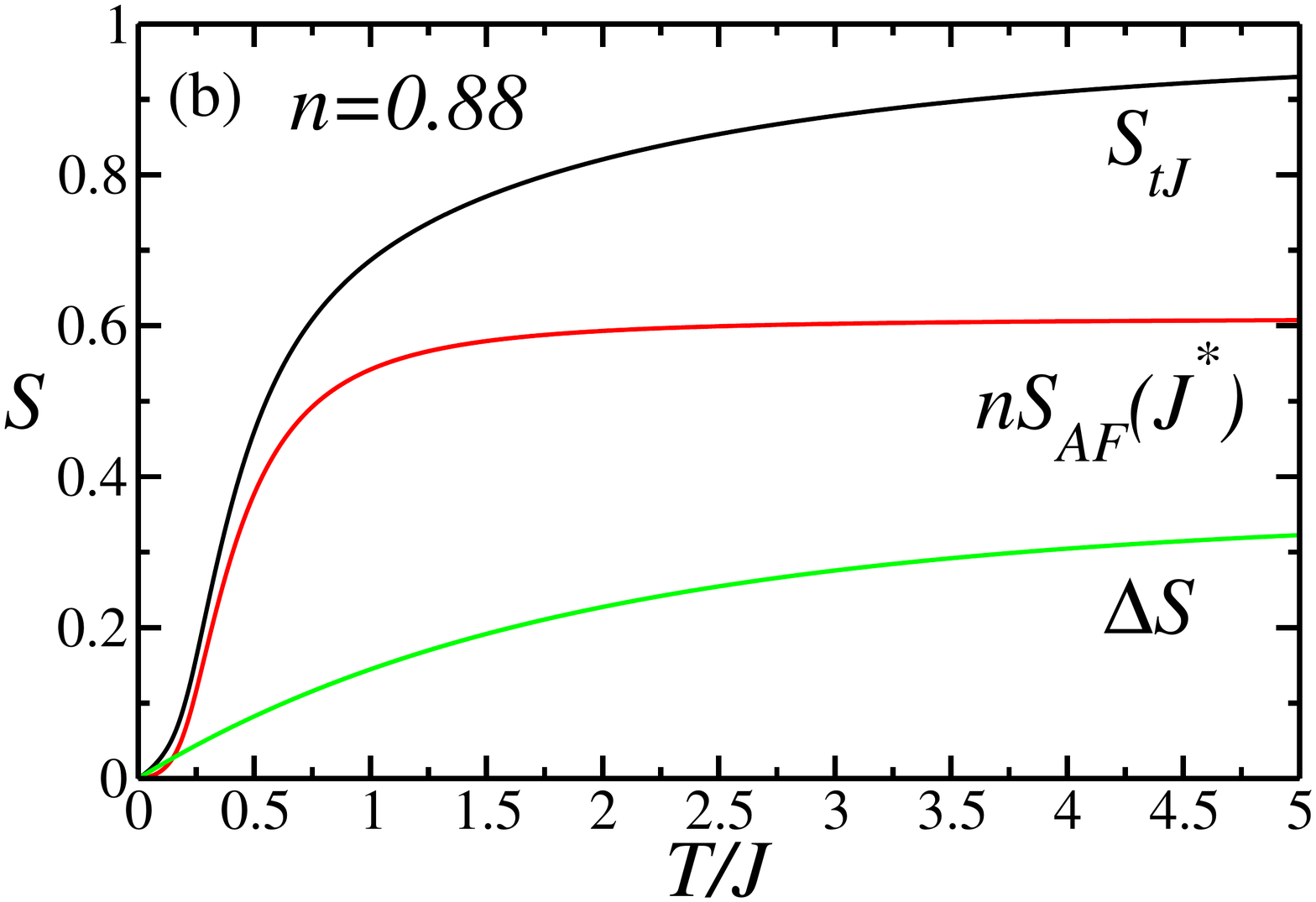}}
 \end{center}
\caption{a) Heisenberg model entropy vs. temperature.  b) An example of the fitting procedure described in the text
         for the $t$-$J$ model entropy at $n=0.88$.  The $t$-$J$ entropy is the sum $S_{tJ}=n S_{AF}(J^*) +\Delta S$,
         with $J^*=0.638173J$.}
\end{figure}
To extract $S_{tJ}$ we need to know $S_{AF}(T)$ which is determined by using a combination of high precision
Monte Carlo data\cite{sandvik1,sandvik2} and series expansion results for $E_{AF}(T)$.  The entropy is found from the energy
by using the thermodynamic relations $E=F+T S$ and $F_{AF}(T)=E_{0AF}-\int_0^T S_{AF}(T')dT'$.  The integal is evaluated by
the trapezoidal rule with step size $\Delta T=10^{-5}J$ and the temperature dependence of the energy is given by
$\tilde{E}_{AF}(T)=E_{AF}(T)-E_{0AF}$ with ground state energy $E_{0AF}=-0.669437J$\cite{sandvik3}.  Using these expressions the
temperature dependence of the entropy $S_{AF}(T)$ is determined iteratively by
\begin{equation}\label{eq:afent}
S_{AF}(j\Delta T) = \frac{2\tilde{E}_{AF}(j\Delta T)}{(2j-1)\Delta T} + \frac{2}{(2j-1)}\sum_{i=1}^{j-1}S_{AF}(i\Delta T),
\end{equation}
where $j$ runs from $1$ to $500,000$ to cover the range $0\le T\le 5J$ and $\tilde{E}_{AF}(0)=S_{AF}(0)=0$.  
Fig(1a) shows the result for the temperature dependence of $S_{AF}(T)$.  Using the relation above between $F_{AF}(T)$, $S_{AF}(T)$
and the value $F_{AF}(5J)=-3.5043614J$ determined from the series expansion the ground state energy found by integrating 
$S_{AF}(T)$ given by Eq(\ref{eq:afent}) is $E_{0AF}^{est}=-0.66947J$ and $E_{0AF}^{est}-E_{0AF}=-3.3\times10^{-5}J$, comparable
to the error estimates in the Monte Carlo data\cite{sandvik1,sandvik2}.
\begin{figure}[h]
  \begin{center}
   \subfigure{\includegraphics[width=0.49\textwidth]{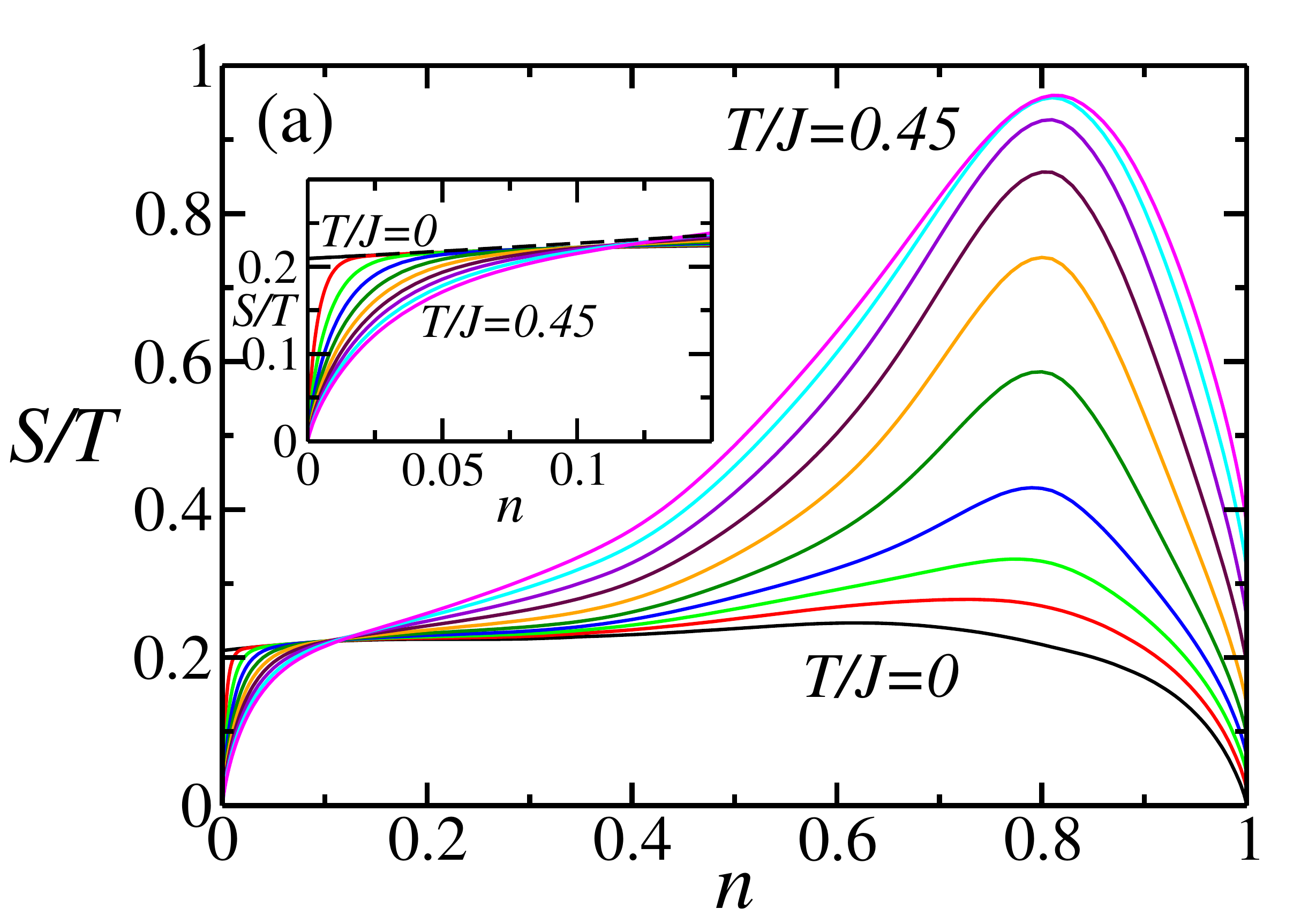}}
   \subfigure{\includegraphics[width=0.49\textwidth]{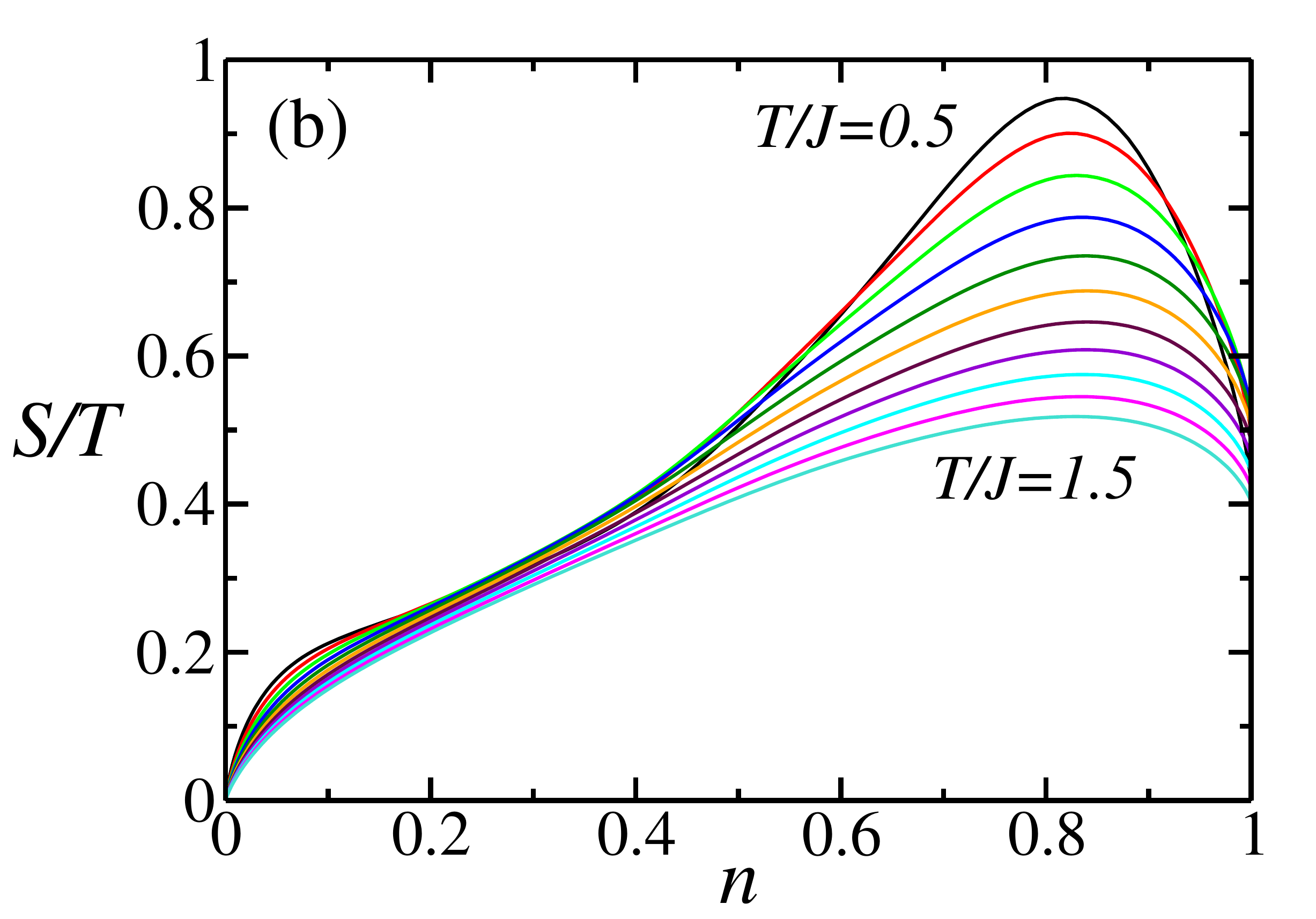}}\\
   \subfigure{\includegraphics[width=0.49\textwidth]{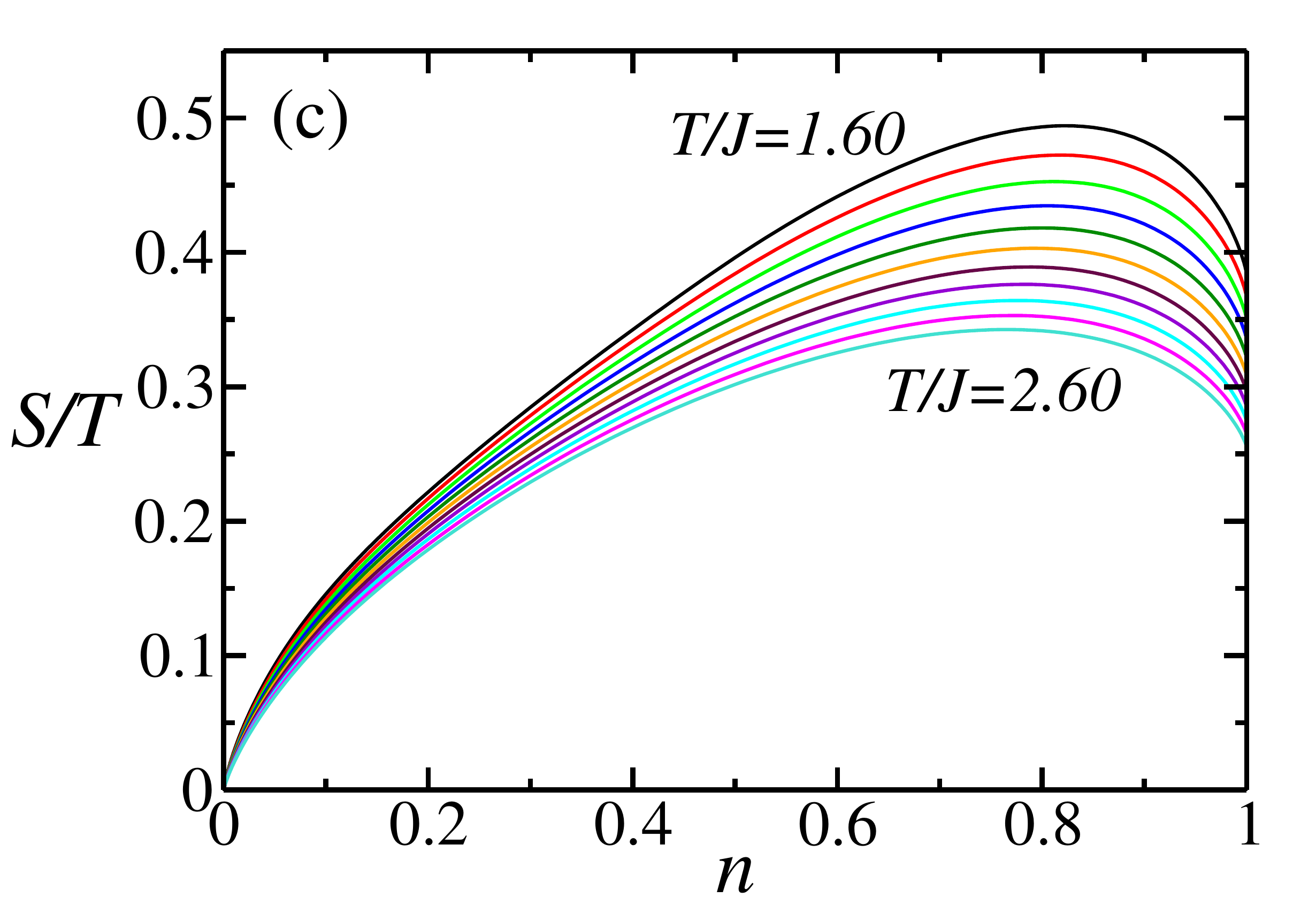}}
   \subfigure{\includegraphics[width=0.49\textwidth]{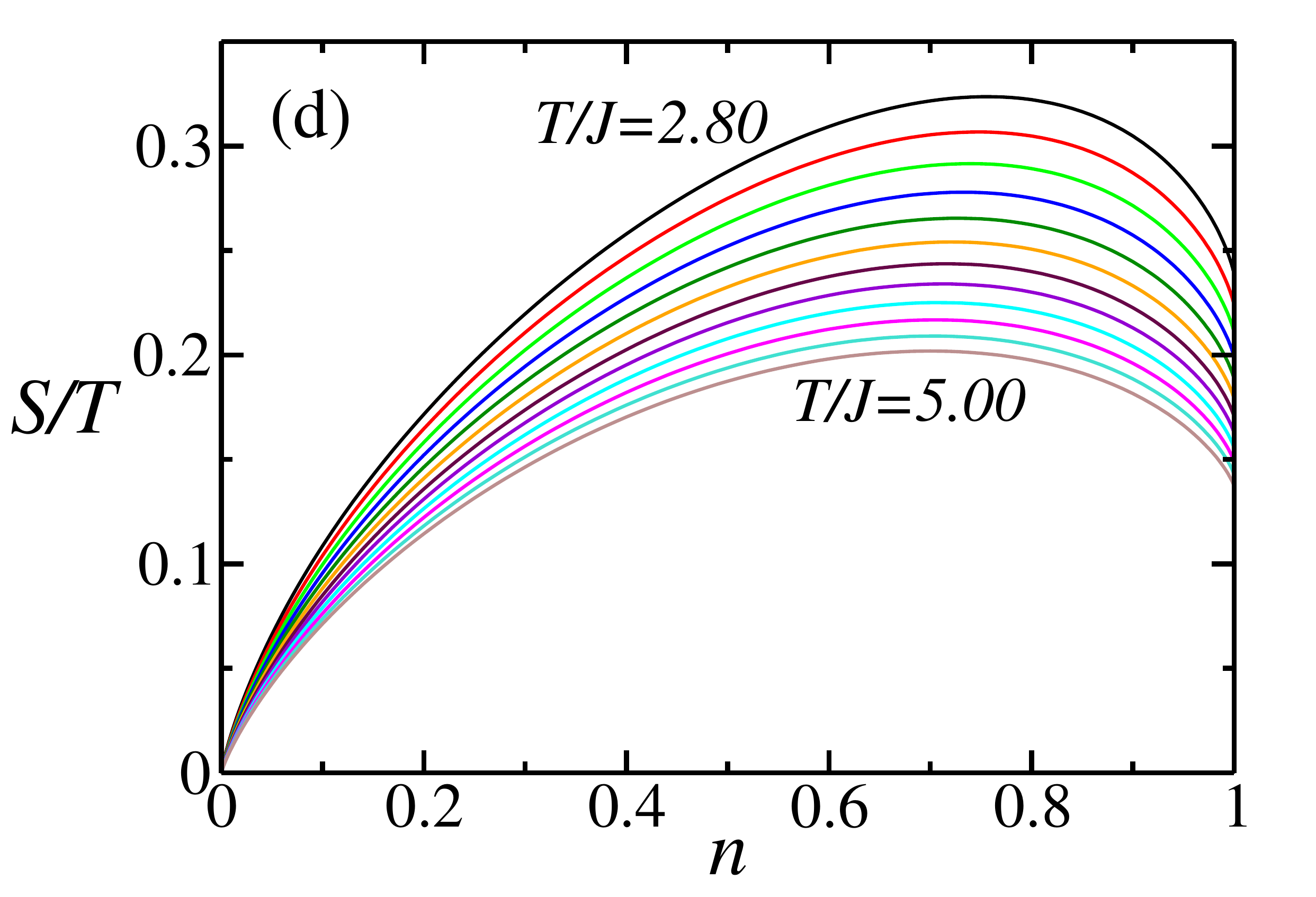}}
  \end{center}
\caption{A selection of data plotted as $S/T$ vs. $n$.  a) $0\le T/J\le 0.45$ in steps of $0.05$.  The dashed line in the
         inset is the tight-binding model $\gamma$.  b) $0.5\le T/J\le 1.5$ in steps of $0.1$.  c) $1.6\le T/J\le 2.6$ in
         steps of $0.1$.  d) $2.8\le T/J\le 5$ in steps of $0.2$.}
\end{figure}
 
The temperature dependence of the Heisenberg entropy $S_{AF}(T)$ can
be used as a known function to reverse the $\Delta S$ calculation and extract $S_{tJ}(n,T)$.
Fig(1b) gives an example of this procedure for $n=0.88$.  Two temperature scales emerge from this analysis: a high 
temperature scale $T\sim t$ and a low temperature scale $T\sim J^*$.  For  $n<0.55$
the $t$-$J$ entropy crosses over to the tight-binding model entropy and all the starred parameters in Eq(\ref{eq:deltas}) are 
required.  A complete discussion of this more complicated fitting procedure at low $n$ is deferred to a future publication,
though here we note to order $n$ the series coefficients for the $t$-$J$ model free energy are exactly the same as the tight-binding
model free energy so as $n\rightarrow 0$ we have $S_{tJ}\rightarrow S_{TB}$ at all temperatures.  

\section{Results}

The extrapolation in temperature to find $S_{tJ}(n,T)$ is done separately for each $n$.  For $0.10\le n \le 0.92$ the 
spacing of densities is $\Delta n=0.01$, while tighter spacing is used for the highest and lowest density ranges.  Data
were accumulated for $226$ densities, with $5,000$ temperature values for each density with a uniform spacing of 
$\Delta T = 0.001 J$.  A selection of the entropy data is shown in Fig(2), with $S_{tJ}/T$ plotted as a function of $n$
for a range of temperatures. 

From the entropy data an interesting quantity to calculate is the density derivative at fixed temperature
$\partial S/\partial n|_T$.  This derivative has been investigated in earlier calculations\cite{prelovsek}, but here we have
greatly improved density resolution allowing a more detailed investigation than before.  The entropy initially grows
upon doping away from half filling and at high temperatures the configurational entropy goes through a maximum at $n=2/3$.
For noninteracting systems the configurational entropy maximum remains true for all temperatures.  For example, the maximum 
entropy for the tight-binding model is found at $n=1$ for all temperatures.  Even for hard core bosons the entropy maximum
remains fixed at $n=0.5$ for all temperatures.  The entropy maximum for the $t$-$J$ model is strongly temperature
dependent as shown in Fig(3).
\begin{figure}[h]
  \begin{center}
  \includegraphics[width=0.9\textwidth]{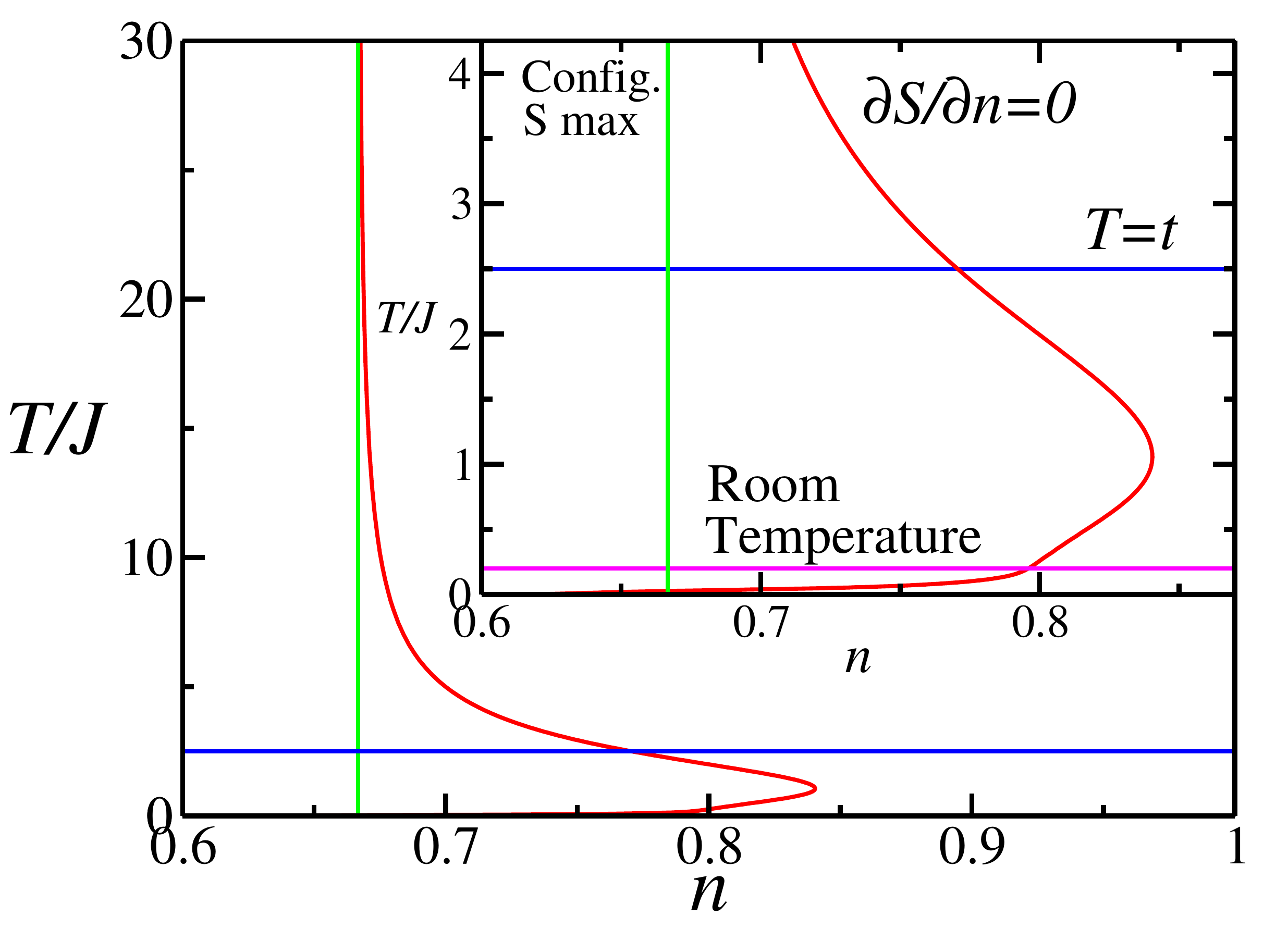}
  \end{center}
\caption{Red line: $\partial S/\partial n=0$ on a temperature vs. density plot.  Green line: configurational entropy
         maximum at $n=2/3$, independent of temperature.  Blue line:  approximate temperature scale $T=t$ for the deviation
         of $\partial S/\partial n=0$ from the configurational entropy maximum.  Magenta line: $T=0.2J$, approximately
         room temperature where $\partial S/\partial n=0$ at $n=0.796$.}
\end{figure}
As the temperature decreases from $T\sim 5J$ to $T\sim J$ the entropy maximum moves up to $n=0.84$.  In this 
temperature range the interactions in the $t$-$J$ model are decreasing the entropy
at higher densities than would be expected from the configurational entropy.  This is an indication of correlations
developing in this density and temperature range.  The temperatures are too high for these to be antiferromagnetic
correlations and the density and temperature ranges don't match the pseudogap found in high temperature superconductors.
For $T\lesssim J$ the entropy maximum reverses and moves back to lower densities, with the temperature dependence decreasing
sharply down to $n=0.8$ and a low temperature tail below $T=0.2J$ extending down to $n=0.62$.  Shastry\cite{shastry} has shown that
$\partial S/\partial n$ is an approximation for the thermopower.  In experiments\cite{obertelli,honma} on cuprate 
superconductors the thermopower
is observed to change sign as a function of doping.  In particular the thermopower is zero at room temperature for
$23$\% doping in many cuprate superconductors\cite{obertelli,honma}.  For $T=0.2J$ (corresponding to room temperature) the $t$-$J$ model entropy 
maximum is at $20.4$\% doping.  

Further information on the correlations developing at lower temperatures in the $t$-$J$ model can be found by considering
the full temperature and density dependence of $\partial S/\partial n$.  By the equality of the mixed partial derivatives
of the free energy we also have
\begin{equation}\label{eq:dmuds}
\left.\frac{\partial\mu}{\partial T}\right|_n = -\left.\frac{\partial S}{\partial n}\right|_T,
\end{equation}
where $\mu$ is the chemical potential.  The relation in Eq(\ref{eq:dmuds}) provides another means to interpret the
derivative.  Fig(4) shows $\partial\mu/\partial T$ as a function of temperature for the range of densities where
$\partial S/\partial n=0$.
\begin{figure}
  \begin{center}
   \includegraphics[width=0.9\textwidth]{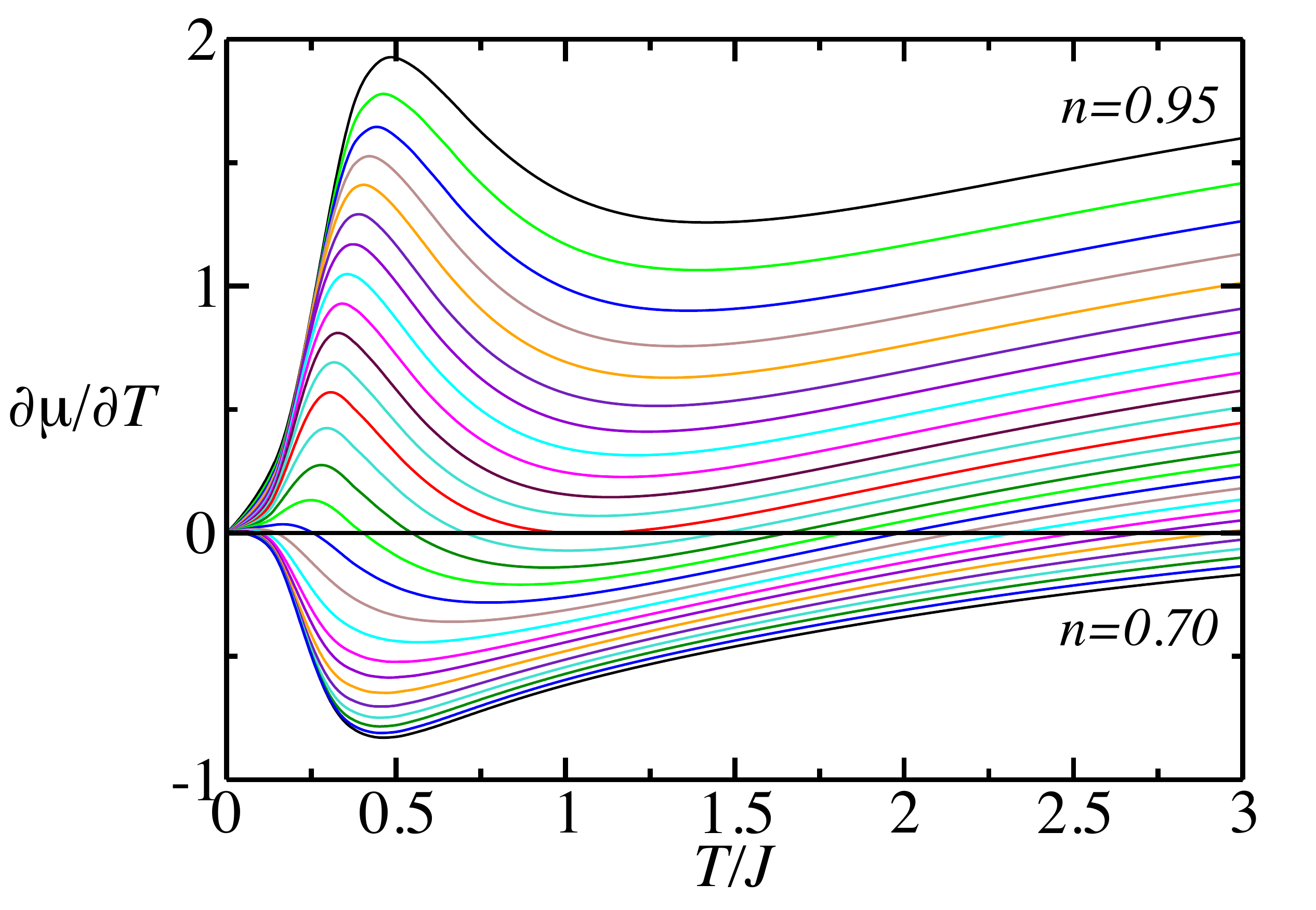}
  \end{center}
\caption{Plot of $\partial\mu/\partial T$ vs. temperature for the density range $0.7\le n\le 0.95$.  The spacing
         of the densities is $0.01$.}
\end{figure}
The low temperature positive peak in $\partial\mu/\partial T$ for $n\gtrsim0.8$ is due to the attractive antiferromagnetic
interaction between oppositely oriented spins decreasing the chemical potential at low temperatures.  It costs less energy 
to add an electron to the system when the temperature falls below the effective spin interaction energy.  For 
$n\lesssim0.8$ in Fig(4) the opposite happens: a fairly broad negative peak in $\partial\mu/\partial T$ develops.  Below
this peak it costs more energy to add an electron to the system than for temperatures above the peak.  For the full range
of densities shown in Fig(4) the high temperature limit of $\partial\mu/\partial T$ is positive.

\section{Conclusions}

The strong temperature dependence of $\partial S/\partial n=0$ shows the interactions in the $t$-$J$ model are producing
competing correlations.  For $n\gtrsim0.84$ antiferromagnetic fluctuations are dominant at low temperatures.  For $n\lesssim0.8$
the dominant fluctuations at low temperature are d-wave pair fluctuations as shown in Ref. \citen{putikka}.  For densities 
$0.8\lesssim n\lesssim 0.84$ there is a crossover between the two dominant fluctuations.  The different fluctuations also
give very different temperature dependences to $\partial\mu/\partial T$.  The antiferromagnetic
fluctuations reduce the energy needed to add an electron at low temperatures while the d-wave pair fluctuations increase
the energy needed to add an electron at low temperatures.

\ack
The author acknowledges the support of the Ohio Supercomputer Center.

\end{document}